\documentclass[aps,prl,twocolumn,groupedaddress]{revtex4}
\usepackage{graphicx}

\begin{document}


\title{Density functional study of the magnetic properties of $Bi_{4}Mn$ clusters: Discrepancy between theory and experiment.}


\author{J. Botana}
\email[]{jorge.botana@usc.es}
\affiliation{Dpto. F\'{i}sica Aplicada, Facultade de F\'{i}sica, Universidade Santiago de Compostela, E-15782, Santiago de Compostela, Spain}
\affiliation{Instituto de Investigaci\'{o}ns Tecnol\'{o}xicas, Universidade de Santiago de Compostela, Santiago de Compostela E-15782, Spain}

\author{M. Pereiro}
\affiliation{Dpto. F\'{i}sica Aplicada, Facultade de F\'{i}sica, Universidade Santiago de Compostela, E-15782, Santiago de Compostela, Spain}
\affiliation{Instituto de Investigaci\'{o}ns Tecnol\'{o}xicas, Universidade de Santiago de Compostela, Santiago de Compostela E-15782, Spain}

\author{D. Baldomir}
\affiliation{Dpto. F\'{i}sica Aplicada, Facultade de F\'{i}sica, Universidade Santiago de Compostela, E-15782, Santiago de Compostela, Spain}
\affiliation{Instituto de Investigaci\'{o}ns Tecnol\'{o}xicas, Universidade de Santiago de Compostela, Santiago de Compostela E-15782, Spain}

\author{J. E. Arias}
\affiliation{Instituto de Investigaci\'{o}ns Tecnol\'{o}xicas, Universidade de Santiago de Compostela, Santiago de Compostela E-15782, Spain}


\date{\today}

\begin{abstract}
We have performed collinear and non collinear calculations on neutral $Bi_{4}Mn$, and collinear ones on ionized $Bi_{4}Mn$ with charges +1 and -1 to find out why theoretical calculations will not predict the magnetic state found in the experiment. We have used the density functional theory to find a fit between the theoretical prediction of the magnetic moment with the experimental value. Our calculations have consisted in a structure search of local energy minima, and then a search of the magnetic lowest energy state for each resulting isomer. The geometry optimization found 3 local minima whose fundamental state is the doublet spin state, which could not be found in previous theoretical works, but they are higher in energy than the lowest-lying isomer by $\approx1.75$ eV. This magnetic state could help understand the experiment. Calculations of non-collinear magnetic states for the $Bi_{4}Mn$ do not lower the total magnetic moment. We conclude arguing how the 3 isomers with doublet state could actually be the ones measured in the experiment.
\end{abstract}


\maketitle

\section{\label{sec1}I. Introduction}
Nanostructured materials have become the main source in current technology to develop devices with novel properties, as low dimensionality causes matter to behave differently than the known properties of bulk. Among them, the zero-dimensional materials, i.e.  atomic clusters and nanoparticles, have attracted a great deal of attention, specially their magnetism. Pure small clusters can exhibit far larger magnetic moment per atom than the isolated atom or their crystals\cite{deheer,alonso,pereiro}, and this effect can be enhanced when impurities are added\cite{wang}. Within this trend, the study of transition metal binary clusters was triggered by the exceptionally high magnetism found in CoRh nanoparticles by Zitoun et al.;\cite{zitoun} since then a number of magnetically enhanced nanoalloys of ferromagnetic and non-magnetic transition metals have been studied theoretically\cite{berlanga,kim,sondon,pereiro2} and experimentally\cite{hihara,pokrant,deheer2}. Yin et al.\cite{deheer3} found an enhanced magnetism in BiMn clusters for Bi-to-Mn ratios close to 2 in their Stern-Gerlach measurements, which deviated from the trend in other Bi nanoalloys like BiCo, where the magnetic moment had a small dependence with cluster size\cite{hihara}. In a later paper, Chen et al.\cite{chen} perform extensive density function theory (DFT) calculations on $Bi_{n}Mn_{m}$ (n=1-6, m=1-12) to learn their structure and how their magnetism works.

While their calculations have been found to be in quite a good agreement with the experiment for this cluster series, there are a few instances of large discrepancies between the theoretical value of the magnetic moment and the measured one. The most significative one, as highlighted by Chen et al., is the $Bi_{4}Mn$ where DFT computation predicts a total magnetic moment,  $\mu_{T}=5 \mu_{B}$ while the experiment measures $1.6\mu_{B}$. Extensive geometric optimization and an estimation of the orbital contribution to the magnetic moment were performed to solve this problem, but the cause of this discrepancy was not found.

Addressing the reason for these disparities is key. One one hand, the existence of an unadressed question in the literature calls for it to be explained. On the other hand, these problems might point out defficiencies in our approach to ab initio calculations, or in the experimental techniques used. The relevance of DFT calculations comes from the reliability of their predictions, and we need to learn the conditions for which they work, and if they do not work for a case, we need to learn the reasons.

In the present paper we will explore the possible sources of the mismatch between theoretical calculations and experimental results for the $Bi_{4}Mn$ nanoalloy, through DFT calculations. The considered possibilities are: finding isomers of the cluster whose total magnetic moment are low enough; allowing non-collinear magnetic states for the studied $Bi_{4}Mn$ structures, as this might lower their magnetic moment; finding out the orbital contribution to the magnetic moment of each structure. We will organize our communication as follows. In Sec. II we explain the computational methods we have used to perform the different types of calculations we performed, also reasoning why each method was used seeking which result. In Sec. III we show our results and discuss them, this section being divided in two subsections: subsec. 1 contains the results of the calculations with imposed collinear magnetic moments, and subsec. 2, the results of the calculations where the collinear is lifted, allowing magnetic moments to arrange freely. Sec. IV contains the conclusions derived from our findings.


\section{\label{sec2}II. Computational Details}
Our first step was to perform ab initio calculations within the framework of the DFT over a large number of different structures of $Bi_{4}Mn$, both taken from the literature and created by the authors. We have solved each system using linear combinations of Gaussian-type orbitals within the Kohn-Sham (KS) density functional methodology (LCGTO-KSDFM), with the deMon 2003 code\cite{koster3}. The calculation of the exchange-correlation (XC) energy term was carried out using the generalized gradient approximation (GGA), in the Perdew-Burke-Ernzerhof ansatz\cite{pbe}. It has been shown that the GGA is the approximation that gives the best results for geometry optimizations of small clusters\cite{pereiro3,yabana}. The orbital basis set used in our calculations\cite{huzinaga} for Bi and Mn uses the effective core potential from Stuttgart-Dresden: RECP|SD\cite{flores}.  It considers a relativistic approximation for the outer 15 electrons of the Mn, and a quasi-relativistic one for the 5 outer electrons of the Bi. The electron density is expanded in auxiliary basis sets in order to avoid the calculation of the $N^4$ scaling Coulomb repulsion energy, where N is the number of the basis functions. The auxiliary basis sets are GEN-A3*\cite{koster} for the Bi and A2-DZVP \cite{koster2} for Mn.
Each structure was geometrically optimized through a Born-Oppenheimer molecular dynamics simulation. After this first approach, the geometry of each cluster was optimized further with a Broyden-Fletcher-Goldfarb-Shanno algorithm\cite{schlegel}. Consecutively, we carried out a search in the collinear magnetic state that yields the energy minimum for zero external magnetic field and T=0 for each geometry.
After this optimization, an initial structure to which we impose different magnetic configurations often relax to different final structures, while different initial structures might converge to the same final structure. Henceforth we obtain a large number of relaxed isomers; we focus on the 21 geometries with lower energy. At this point we perform single-point calculations for each isomer, searching for their fundamental magnetic state.
Next we proceed to perform non-collinear single-point calculations to determine the lowest-energy magnetic state for each one of the 21 isomers. These calculations are performed using a different DFT 
implementation, the Octopus code\cite{marques2}. The Octopus calculations have considered wavefunctions to be complex spinors and they include the spin-orbit coupling relativistic correction. In the same manner as the collinear calculations, XC energy is approximated through a GGA-PBE functional. The inner electrons of the atoms are represented with a core model, using the Hartwigsen-Goedecker-Hutter pseudopotentials.\cite{hgh} As for the grid upon which the KS equations are solved, we have chosen a radius of 4.4 and a spacing of 0.1. The mixing scheme is a broyden one, with a mixing factor of 0.02.

\section{\label{sec3}III. Results}
\subsection{\label{col}1. Collinear Calculation Results}
\begin{figure}
	\centering
		\includegraphics[width=1.0\textwidth]{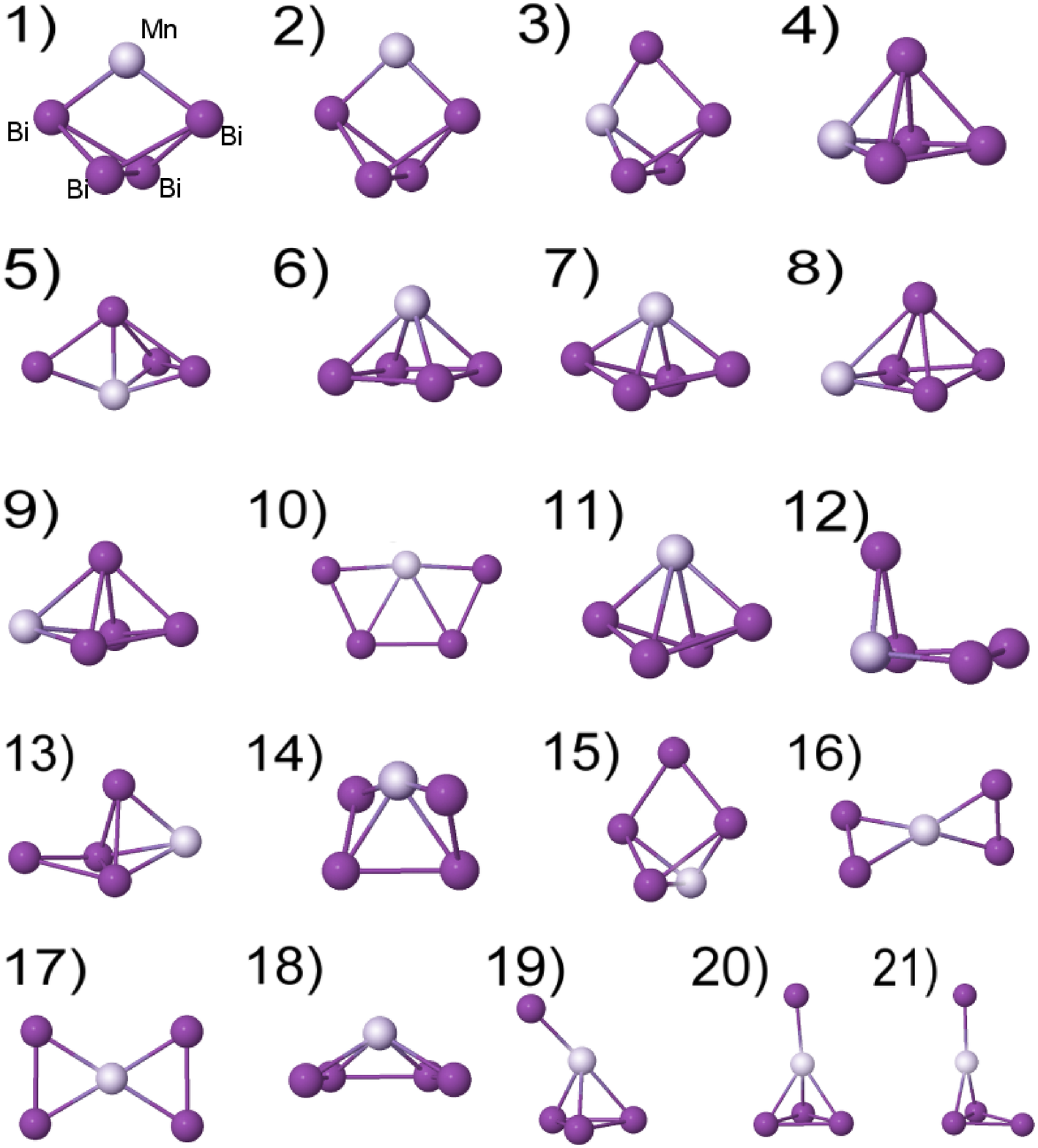}
	\caption{(color online) Geometry of the 21 lowest energy structures for the nanoalloy $Bi_{4}Mn$, from 1) lowest to 21) higher.}
	\label{dMstructures}
\end{figure}

\begin{figure}
	\centering
		\includegraphics[width=1.0\textwidth]{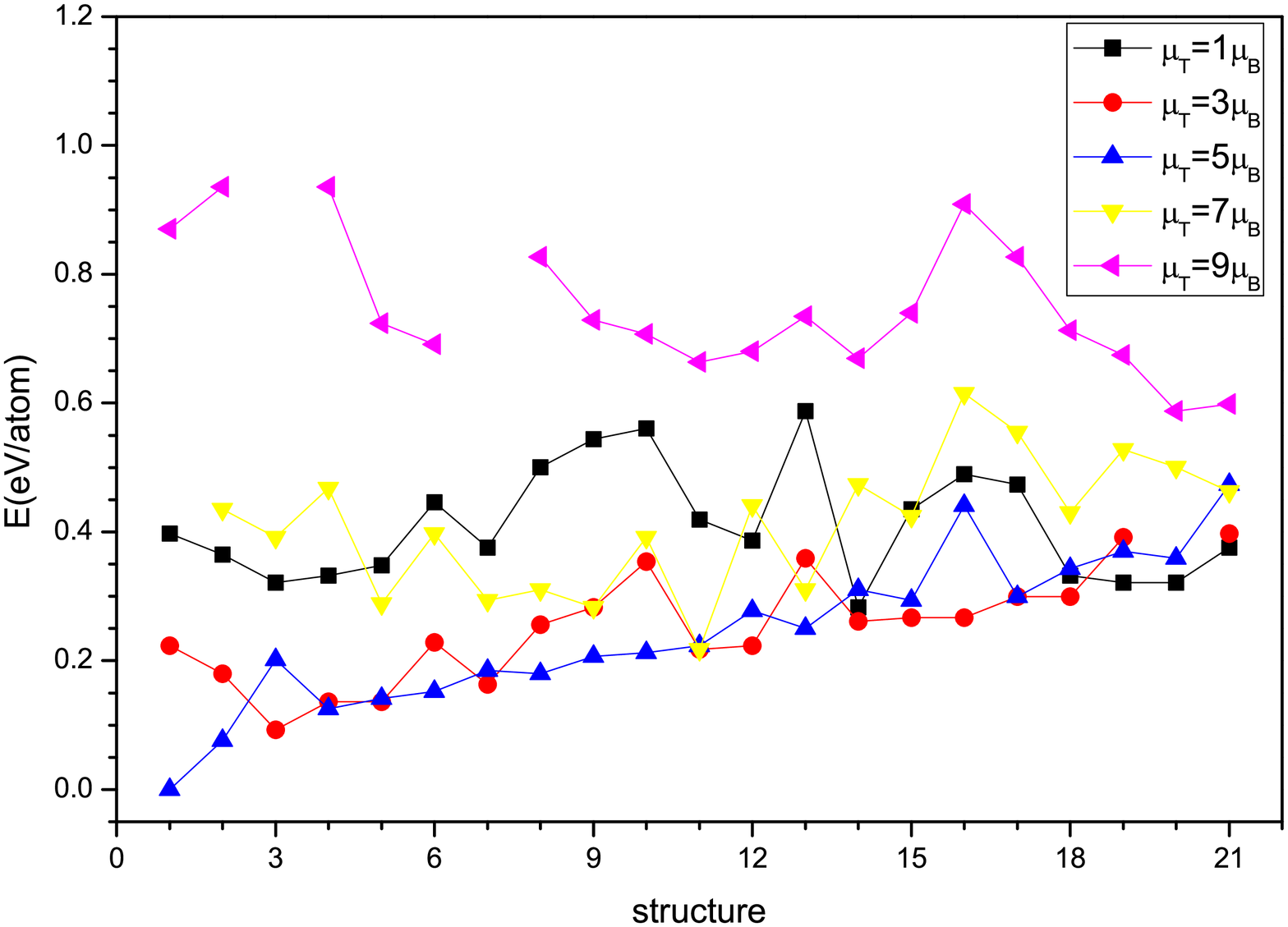}
	\caption{(color online) Energy comparison of the 21 structures with lower energy for the magnetic configurations $\mu_{total}=1,3,5,7,9\mu_{B} $. The numbers in the x-axis correspond to the structures shown in Fig. \ref{dMstructures}.}
	\label{dMenergies}
\end{figure}

In Fig. \ref{dMstructures} we represent the 21 lowest-energy isomers that we have found in our calculations and in Fig. \ref{dMenergies} we plot the energies of each of their magnetic states from 1 to $9 \mu_{B}$. Our lowest energy structure, an edge-capped distorted tetrahedron, and its $\mu_{total}=5 \mu_{B}$ agree with the previous theoretical calculations, which in turn disagree with the experimental data\cite{deheer3}. Also, we have found the pyramidal structure to be 0.76 eV higher than the lowest lying isomer, and the W-like structure 1.06 eV higher, which is in very good agreement with Chen's work.\cite{chen} Furthermore, isomers with low coordination for the Mn tend to have lower energies as well.

The structures with $\mu_{total}=5 \mu_{B}$ dominate the lowest energy isomers until the $11^{th}$ one, and from that point on, structures with $\mu_{total}=3 \mu_{B}$ dominate instead until the the $18^{th}$ one. Our 3 highest energy isomers have $\mu_{total}=1 \mu_{B}$. Studying separately the different isomers for each of these 3 magnetic states, we have found that the energy of clusters set to $\mu_{total}=3 \mu_{B}$ and $5 \mu_{B}$ grows in average with the ground state energy. On the other hand, if the clusters are kept at $\mu_{total}=1 \mu_{B}$, we find no such tendency.

\begin{figure}
	\centering
		\includegraphics[width=1.0\textwidth]{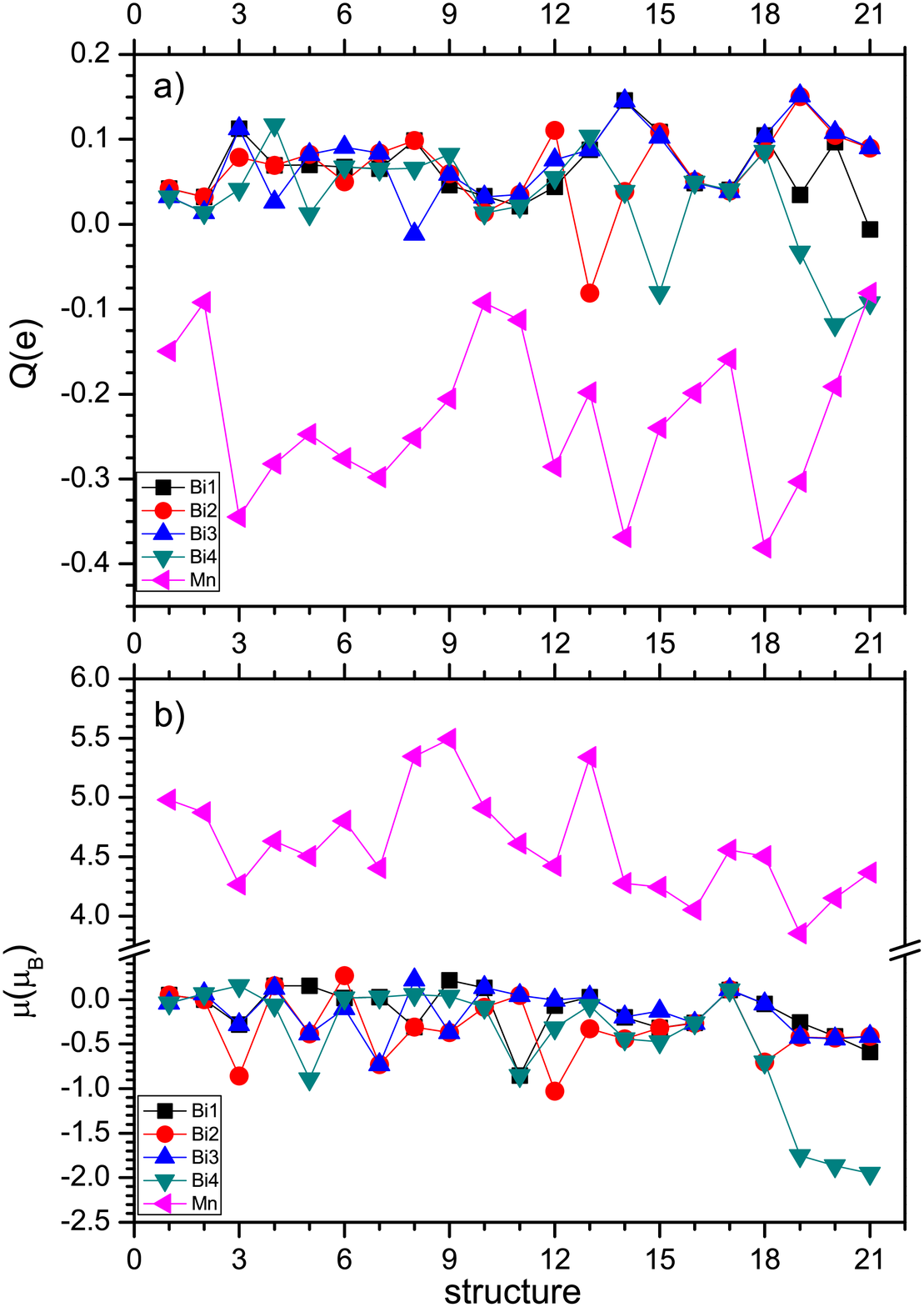}
	\caption{(color online) Atomic charges (a) and magnetic moments (b) of each atom of the ground state of each one of the 21 lowest energy isomers of $Bi_{4}Mn$.}
	\label{charge}
\end{figure}

Analysing each atom of the nanoalloy separately, most of the magnetic moment comes from the Mn atom, while the Bi magnetic moments tipically remain close to zero. As we can see in Fig. \ref{charge}a), electrons are shared from the Bi to the Mn, in particular from $6p$ orbitals to $4d$. This transfer does not generally enhance magnetism in either element, as the Mn total magnetic moment oscillates between $4$ and $5 \mu_{B}$ for all clusters, without correlation to charge transfers (see Fig. \ref{charge}b)). In the 3 higher energy isomers, however, one of the Bi atoms stands isolated, bound only to the central Mn. This Bi, in addition to transferring $6p$ electrons to the Mn, also receives charge into its $7p$ orbital, greatly increasing its magnetic moment which couples antiparallel with the Mn. This makes the total magnetic moment of the isomer to decrease to $1 \mu_{B}$.
The lower total magnetic moment state is hence originated in the enhancement of the magnetism of the Bi atoms, whose magnetic moments are either very close to zero or antiparallely aligned to those of the Mn atoms. 

\begin{figure}
	\centering
		\includegraphics[width=1.0\textwidth]{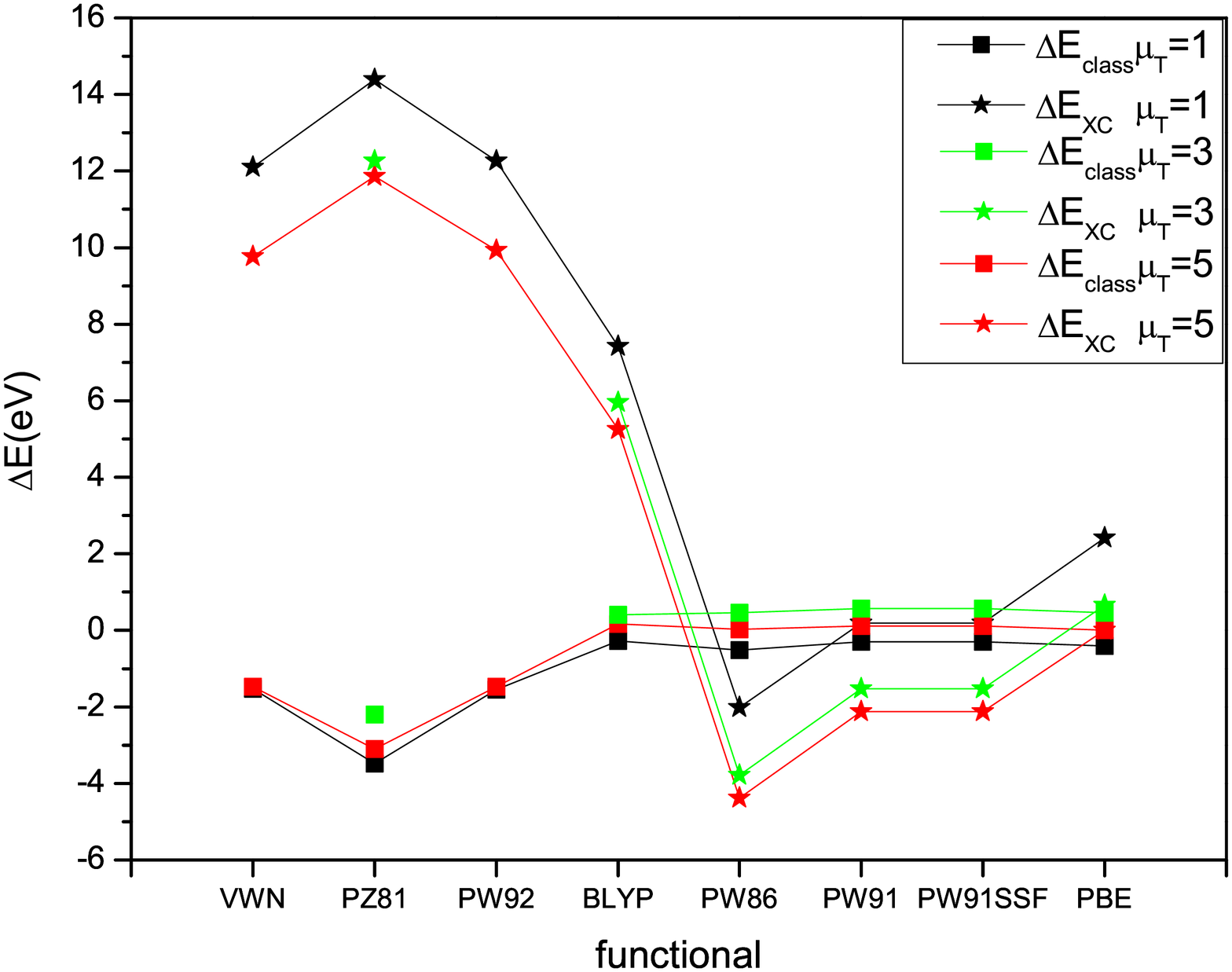}
	\caption{(color online) Comparison of the 3 lowest magnetic states of the lowest lying isomer of $Bi_{4}Mn$ calculated using different XC functionals.}
	\label{functionals}
\end{figure}

We have not found the lowest energy isomer to have a total magnetic moment close to the experimental value. We have found the structures whose ground state is close to the experimental total magnetic moment to be energetically close enough to the lowest energy one to have a large enough population in the synthesis to justify the experimental results. So we proceed to test other exchange-correlation functionals to find out if this could be the reason for the discrepancy. In Fig. \ref{functionals} we show the ground states calculated for the lowest-lying isomer for different exchange-correlation (XC) functionals, and we compare them with our results for PBE functional. We performed this calculation for the values of $\mu_{T}=1,3,5 \mu_{B}$. Local functionals exhibit higher energies, but all functionals provide the same energetic ordering of the three magnetic states: $\mu_{T}=5 \mu_{B}$ as the lowest, $3 \mu_{B}$ second lower and $1 \mu_{B}$ third lower. When splitting the energy in XC and classical terms, all functionals keep the same ordering: XC term greatly favors the $3 \mu_{B}$ and $5 \mu_{B}$ states, while the bias of the classical term lowering the $1 \mu_{B}$ state cannot balance it.

\begin{figure}
	\centering
		\includegraphics[width=1.0\textwidth]{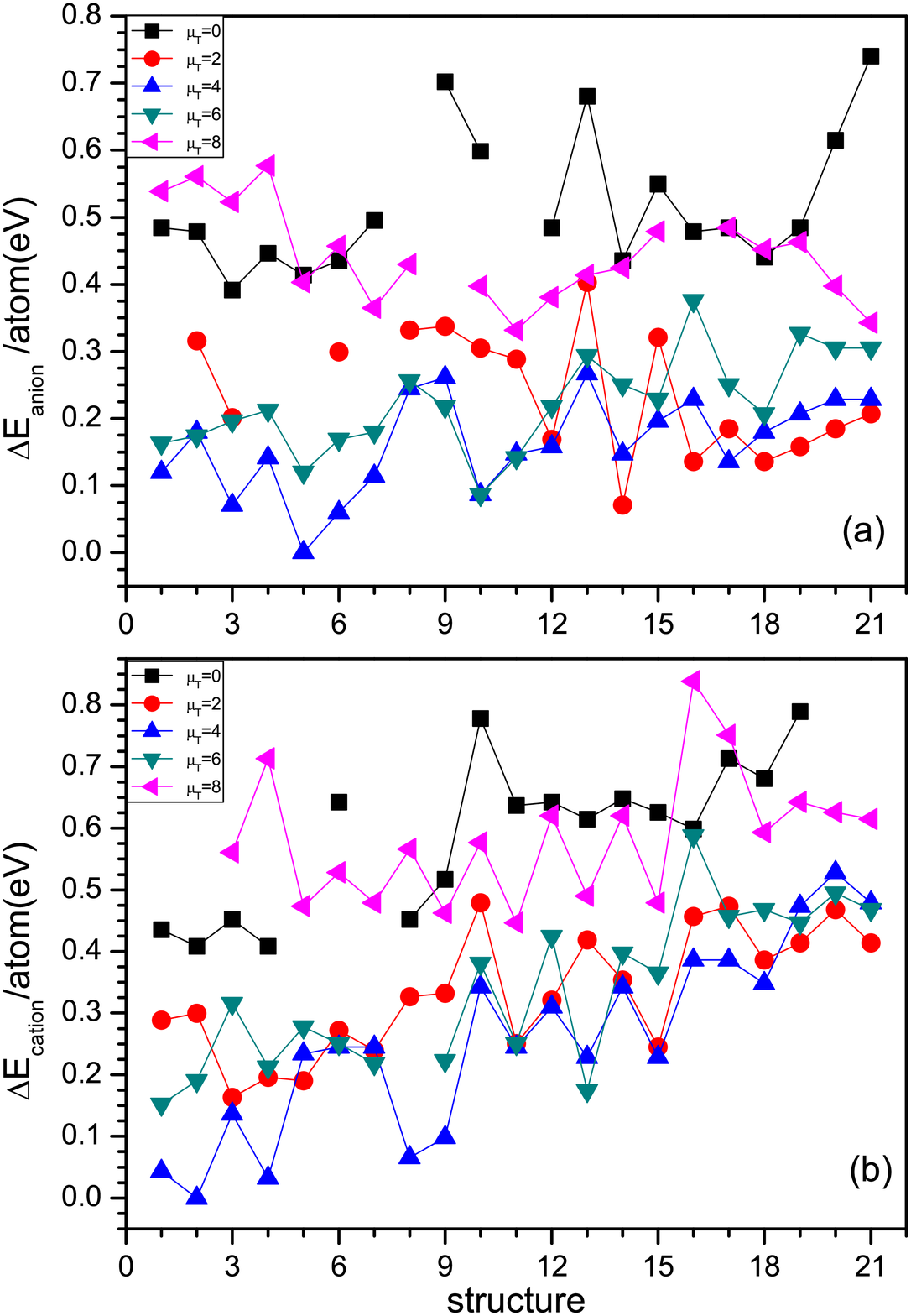}
	\caption{(color online)Energy comparison of the 21 structures with lower energy for the magnetic configurations $\mu_{T}= 0, 2, 4, 6, 8 \mu_{B}$ of the negatively (a) and positively (b) charged ions of $Bi_{4}Mn$.}
	\label{ionenergies}
\end{figure}

The final scenario we can examine with collinear calculations is the possibility that the experimental BiMn clusters are ionized in the moment of their production, or that their structure changes during the ionization in Stern-Gerlach. This is unlikely because of the low temperature of the process.\cite{moro} The results still do not yield any significative reduction of the total magnetic moment, as we can see in Fig. \ref{ionenergies}. The lowest lying isomers are small deformations of the one for the neutral cluster, and the lowest magnetic configuration is for either ion $4\mu_{B}$. The lowest magnetic state is found for the anion third lowest lying isomer, which has $\mu_{T}=2\mu_{B}$. Even this result is larger than the experimental value, so it cannot explain the discrepancy.

As a summary, we have found that there are 3 structures of the $Bi_{4}Mn$ cluster whose $\mu_{T}=1 \mu_{B}$. The experimental value is $1.6 \mu_{B}$, which implies that it is not isomerically pure: there has to be a population mixture of clusters with different structures and different $\mu_{T}$, including $1 \mu_{B}$. But these 3 structures are $\approx1.75 eV$ higher than the lowest-lying one, and there are 18 structures with lower energy than them. Consequently, it is unlikely that the experimental device can produce these isomers in a high enough ratio as to lower the magnetic moment, explaining the experimental value. Except that some other reason, unaccounted for in our calculations, happens to favour their synthesis.

\subsection{\label{noncol}2. Non-Collinear Calculation Results}
\begin{figure}
	\centering
		\includegraphics[width=1.0\textwidth]{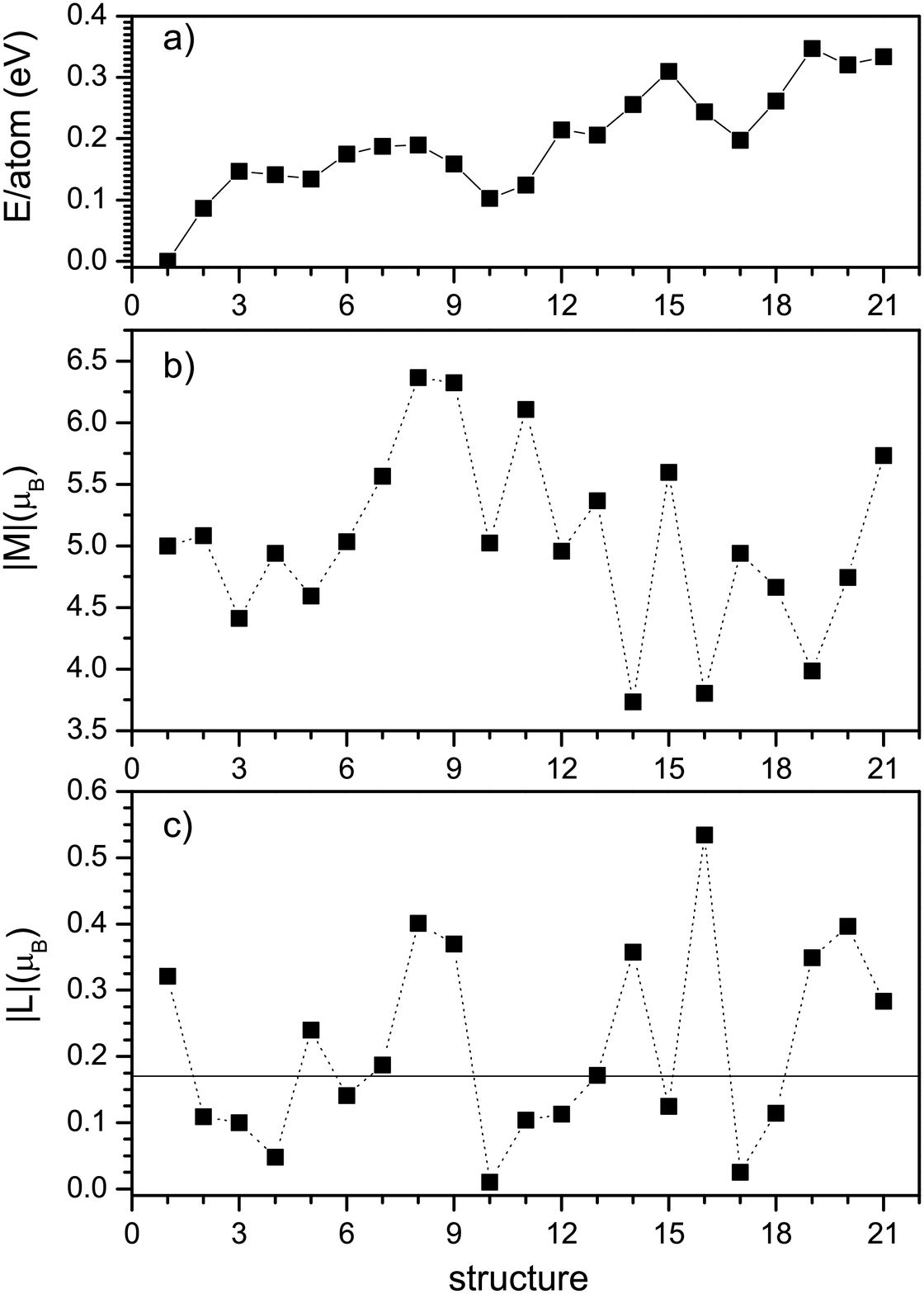}
	\caption{(color online)Energy (a), magnetic moments (b) and orbital moments (c) for the 21 lowest energy isomer of $Bi_{4}Mn$ found in the collinear calculations. In c), the horizontal line marks the value of the orbital magnetic moment in bulk BiMn.}
	\label{noncolfig}
\end{figure}

The next step is finding out if the discrepancy can be solved with non-collinear calculations, and we perform such calculations for the 21 lowest-energy structures we found in the collinear calculations. In table \ref{noncoltbl} we show the results, with the energy and the components of the total magnetic moment for each geometry. As we can see in Fig. \ref{noncolfig}a), the energy dependence of the different isomers with non-collinear magnetic configuration roughly follows that of the isomers with collinear magnetism. The total magnetic moment (see Fig. \ref{noncolfig}b)) ranges from $3.7$ to $6.4 \mu_{B}$, with the lowest lying isomer having $\mu_{T}=5.0 \mu_{B}$, values which exceed the experimental value and agree with the collinear calculations for the lowest energy clusters. In the last column of the Table \ref{noncoltbl} we compare the modulus of the total magnetic moment components: $M_{x}$ and $M_{y}$ against $M_{z}$. This ratio is very small, specially for the lower energy isomers, hence we can conclude that the collinear approximation is a good one, which validates the results of our collinear calculations in the previous subsection.

Further analysis of the magnetic components of our isomers yield the orbital magnetic moment: In Fig. \ref{noncolfig}c) we compare the modulus of the orbital magnetic moment with the value for BiMn in bulk, $0.17 \mu_{B}$,\cite{coehoorn} seeing that the value of the orbital magnetic moment has the same order of magnitude, being small in all cases. In the column 5 of Table \ref{noncoltbl} we compare the orbital component with the total magnetic moment of each isomer, and we find that in all cases, this contribution is smaller than $10\%$, except for the structure 16, where the ratio between the moduli is $14\%$. From our results, we conclude that the orbital magnetic moment for these clusters is too small to be a significative factor in the total magnetic moment, so it cannot be the source of the discrepancy between the experimental results and the theoretical calculations. Furthermore, our calculated orbital magnetic moments for each isomer are well within the upper limit estimated by Chen et al., $1.35 \mu_{B}$.

\begin{table}
	\centering
		\begin{tabular}{c c c c c c}
			\hline
			\hline
struc &	$^E/_{atom}$ &	$|M|$ &	$|L|$ &	$^{|L|}/_{|M|}$ &	$^{M_{xy}}/_{M_{z}}$\\
		\hline
 1 & 0.000 &  5.0 & 0.321 & 0.060 & 0.00\\
 2 & 0.087 &  5.08 & 0.109 & 0.02 & 0.00\\
 3 & 0.147 &  4.41 & 0.100 & 0.02 & 0.02\\
 4 & 0.141 & 4.94 & 0.048 & 0.01 & 0.00\\
 5 & 0.134 &  4.6 & 0.240 & 0.05 & 0.10\\
 6 & 0.175 & 5.03 & 0.141 & 0.03 & 0.01\\
 7 & 0.188 &  5.57 & 0.187 & 0.03 & 0.12\\
 8 & 0.190 &  6.37 & 0.401 & 0.06 & 0.04\\
 9 & 0.159 &  6.32 & 0.369 & 0.06 & 0.02\\
10 & 0.103 & 5.03 & 0.010 & 0.002 & 0.01\\
11 & 0.124 &  6.11 & 0.104 & 0.02 & 0.14\\
12 & 0.214 & 4.96 & 0.113 & 0.02 & 0.01\\
13 & 0.206 & 5.37 & 0.171 & 0.03 & 0.04\\
14 & 0.256 & 3.73 & 0.358 & 0.10 & 0.05\\
15 & 0.310 &  5.59 & 0.125 & 0.02 & 0.02\\
16 & 0.244 & 3.8 & 0.534 & 0.14 & 0.06\\
17 & 0.197 & 4.94 & 0.025 & 0.01 & 0.01\\
18 & 0.262 &  4.66 & 0.114 & 0.03 & 0.11\\
19 & 0.347 & 3.99 & 0.349 & 0.09 & 0.12\\
20 & 0.320 & 4.74 & 0.397 & 0.08 & 0.05\\
21 & 0.334 & 5.73 & 0.283 & 0.05 & 0.16\\
			\hline	
			\hline
		\end{tabular}
	\caption{\label{noncoltbl}Spread of the results obtained in our non-collinear calculations. First column labels each structure according to their energy ordering in the collinear calculations. Second column units are electronvolts. Column 3 is the modulus of the total magnetic moment and units are $\mu_{B}$.Column 4 is the modulus of the orbital moment and units are $\mu_{B}$. Column 5 is the dimensionless ratio of the orbital moment and the total magnetic moment. Column 6 is the dimensionless ratio of the XY plane component of the total magnetic moment and the Z axis component.}
	
\end{table}

\section{\label{sec4}IV. Conclusions}
We have explored the possible scenarios that could lead to the known discrepancy between the experimental value of the total magnetic moment of the $Bi_{4}Mn$ cluster and the theoretically calculated one. We have found that three isomers among the 21 most stable actually have a total magnetic moment below the experimental one, but these isomers are too high in energy respect to the lowest lying one. Furthermore, there are 18 isomers with lower energy, so these three cannot make up for a fraction of the population of randomly created $Bi_{4}Mn$ clusters significative enough to reduce the average total magnetic moment down to the experimental value. We have not found either that using different XC functional approximations makes the ground state of the lowest lying isomer have a lower magnetic moment. The analysis of the positive and negative singly-charged ions has also yielded no lowest energy structure with a magnetic moment closer to the experimental one. In fact, none of the ions of the 21 isomers we have considered has a ground state with a magnetic moment lower than the experiment. This should be expected, though, as the only lower magnetic state available to ions of $Bi_{4}Mn$ is $0 \mu_{B}$. The possible presence of ions would not help us explain the experiment unless we had found ground states at $\mu_{T}=0 \mu_{B}$.
From our non-collinear calculation, we rule out other two possible sources of the disagreement: the magnetic configuration is collinear to a high degree, even when allowing the individual atomic magnetic moment to arrange freely, and still produces states with high magnetic moment: between $4$ and $6.4 \mu_{B}$, hence this cannot be the source; also, the orbital magnetic moment is too low in absolute value compared to the total magnetic moment to produce a significant reduction in it.

All these consistently negative results in our search for a source of the discrepancy in the calculations suggest that said source is not an actual error of the calculations. Furthermore, the possibility that DFT method itself is not reliable to study this cluster is unlikely: the DFT results are in fairly good agreement for almost all the other $Bi_{m}Mn_{n}$ clusters. Having this in account, we are only left with the explanation suggested in Sec. III.1, i.e. that the experimental sample of $Bi_{4}Mn$ is composed of a population of different structures and that one or some of our structures 19, 20, 21 are a significative fraction of said sample. We have found no reason why DFT would increase the energy of these structures specifically, so there could be a factor in the experiment that favors their production. We think it is remarkable that the structure 19, 20, 21 are very similar the lowest lying isomer for $Bi_{3}Mn$ clusters (a tetrahedron), simply adding an extra Bi atom to the Mn end of the tetrahedron. If $Bi_{3}Mn$ clusters form much more quickly in the experimental device than $Bi_{4}Mn$ ones, it is not unreasonable to think that the later will form from the former. $Bi_{3}Mn$ also has a large electrostatic dipole we have calculated to be 2.4 D in the direction that connects the Mn atom with the center of the triangle formed by the three Bi atoms. This dipolar moment could help a fourth Bi atom to couple to the Mn instead of breaking up the tetrahedron to form the calculated $Bi_{4}Mn$ lowest lying structure. If the process of measurement of mass and magnetic moment of the clusters is fast enough, they might not have enough time to relax into the said lowest lying isomer, hence resulting in some of our structures 19, 20, 21 making up a large fraction of the measured $Bi_{4}Mn$ clusters.

To verify this, one possible path would be to perform an analysis of the optical properties of the experimentally obtained of the $Bi_{4}Mn$ clusters to identify their geometry, and see if they match with the isomers theoretically obtained as the lowest energy ones, or instead they match with any of the 19, 20, 21 isomers that actually show a low total magnetic moment. If the later case were true, then we would have to explain why the experimental setup produces structures that theoretically at T=0K are known not to be the fundamental one. On the other case, it would be necessary to know why the theory cannot predict the correct magnetic moment for a cluster with known composition and structure.

\section{\label{sec5}V. Acknowledgements}
We wish to acknowledge the Centro de Supercomputaci\'{o}n de Galicia (CESGA) for the computing resources and the advice they have made available for us. We would like to thank Dr. Pardo for helpful discussions. This research has been done under the projects No. MAT2009-08165 and No. INCITE08PXIB236052PR. One of the authors has been enjoying financial support from the Isabel Barreto program.

\end{document}